# Thickness-independent transport channels in topological insulator $Bi_2Se_3$ thin films

Namrata Bansal[1], Yong Seung Kim[2,3], Matthew Brahlek[2], Eliav Edrey[2], and Seongshik Oh[2,*]

[1]Department of Electrical and Computer Engineering, Rutgers, The State University of New Jersey, 94 Brett Rd., Piscataway, NJ 08854, USA

[2]Department of Physics & Astronomy, Rutgers, The State University of New Jersey, 136 Frelinghuysen Rd, Piscataway, NJ 08854, USA

[3]Graphene Research Institute, Sejong University, Seoul 143-747, South Korea

[*]Correspondence: ohsean@physics.rutgers.edu

**Abstract**

**With high quality topological insulator (TI) $Bi_2Se_3$ thin films, we report thickness-independent transport properties over wide thickness ranges. Conductance remained nominally constant as the sample thickness changed from 256 to ~8 QL (QL: quintuple layer, 1 QL ≈ 1 nm). Two surface channels of very different behaviors were identified. The sheet carrier density of one channel remained constant at ~3.0 × $10^{13}$ $cm^{-2}$ down to 2 QL, while the other, which exhibited quantum oscillations, remained constant at ~8 × $10^{12}$ $cm^{-2}$ only down to ~8 QL. The weak antilocalization parameters also exhibited similar thickness-independence. These two channels are most consistent with the topological surface states and the surface accumulation layers, respectively.**

Over the past few years, topological insulators (TIs) have emerged as an ideal platform for spintronics, quantum computations, and other applications [1-9]. They are predicted to have an insulating bulk state and spin-momentum-locked metallic surface states. This spin-momentum-locking mechanism and their band structure topology are predicted to prevent the surface metallic states from being localized. Among the TIs discovered so far, $Bi_2Se_3$ is considered one of the most promising because it has the largest bulk band gap of 0.3 eV and a well-defined single Dirac cone at the momentum zero point in *k*-space [9]. Numerous reports have confirmed the presence of the topological surface states in this material [7-8, 10-15]. However, its bulk state always turns out to be metallic instead of insulating, and so identifying the surface states in transport studies has been challenging. Although one obvious way to suppress the bulk conductance and sort out the surface contribution would be to make the sample thin until the surface contribution dominates, such a simple approach has so far evaded clear answers due to challenging material issues such as thickness- and environment-dependent bulk properties [13, 16]. In this Letter, we report transport properties of a series of high quality $Bi_2Se_3$ thin films taken with well-controlled measurement protocols: we achieved dominant surface transport properties up to a few hundred nanometers in film thickness and identified two surface channels of different origins.

The $Bi_2Se_3$ films used for this study were grown on c-axis $Al_2O_3$ substrates ($10 \times 10 \times 0.5$ $mm^3$) with molecular beam epitaxy (MBE); the films were grown using the recently developed two-step scheme [17]; see Supplemental Material SB [18]. The sharp reflection high energy electron diffraction (RHEED) pattern in Fig. 1(a) exhibits the high crystallinity of the film, and the atomically flat terraces observed by atomic force microscopy (AFM) in Fig. 1(b) are much larger than any previous reports on $Bi_2Se_3$ thin films [11, 17, 19-20], representing the high quality of these samples.

On these samples, transport measurements were made within 20 minutes of the sample being taken out of the MBE chamber in order to minimize the atmospheric doping effect [16]; see Supplemental Material SA [18] for measurement details. Figure 2(a) shows that the resistance vs. temperature (from 290 K to 1.5 K) dependence is metallic down to ~30 K for all thicknesses (2 – 256 QL). Below ~30 K the resistance remained almost constant, indicating static disorders as the dominant scattering mechanism, except for ultrathin films, which show slight resistance increase as temperature decreases. The first notable feature in Fig. 2(a) is that the low temperature resistance is quite thickness independent for samples between ~8 and 256 QL in thickness. This can be seen more clearly in the plot of conductance ($G_{xx}$) at 1.5 K versus sample thickness in Fig. 2(b). Within small error bars, $G_{xx}$ is nominally constant for samples between ~8 and 256 QL thick. This observation suggests that the conductance in this thickness range is dominated by some surface transport channels.

Hall effect measurement shown in Fig. 2(c) provides more insights regarding the origin of these surface channels. If all carriers had the same mobility, $R_{xy}(B)$ should appear as a straight line with the slope determined by $1/(n_{SC}e)$, with $n_{SC}$ representing the total sheet carrier density. However, if there are multiple types of carriers with different but comparable mobilities, nonlinearty shows up in the $R_{xy}(B)$ data; those carriers that have orders of magnitude lower carrier densities or mobilities than others do not affect $R_{xy}(B)$. Therefore, the nonlinearity in Fig. 2(c) suggests the presence of multiple carrier types with comparable mobilities. Specifically, if two carrier types dominate the Hall effect, $R_{xy}(B)$ is given by $R_{xy}(B) = -(B/e)[(n_1\mu_1^2+n_2\mu_2^2)+B^2\mu_1^2\mu_2^2(n_1+n_2)][(n_1\mu_1+n_2\mu_2)^2+B^2\mu_1^2\mu_2^2(n_1+n_2)^2]^{-1}$, where $n_1$ and $n_2$ represent the two sheet carrier densities, $\mu_1$ and $\mu_2$ represent their respective mobilities, e is the electron charge, and B is the magnetic field. It turns out that this two carrier model nicely fits all our Hall resistance data as shown in Fig. 2(c). This implies that the mobilities of all the significant conductance channels in our samples can be approximately grouped into two; in this model, carriers on opposite surfaces or on different bands will appear as part of the same channel if they have similar mobilities. In order to maximize the fitting reliability, we used the Hall conductance, $G_{xy}(B) \equiv -R_{xy}/(R_{xy}^2+R_{xx}^2)$ instead of $R_{xy}(B)$ for the fitting and also reduced the number of fitting parameters to two by applying extra confinement from $G_{xx}(B)$; for details, see Supplemental Material SC [18]. From this two parameter fitting, we extracted the four quantities, $n_1$, $n_2$, $\mu_1$ and $\mu_2$ for each sample.

The most notable feature in Fig. 2(d) is that one channel ($n_{SC-1}$) provided nearly constant sheet carrier density of $\sim 3.0 \times 10^{13}$ cm$^{-2}$ all the way down to 2 QL, whereas the other channel ($n_{SC-2}$) stayed at $\sim 8 \times 10^{12}$ cm$^{-2}$ down to $\sim 8$ QL but gradually decreased for thinner samples. This observation suggests first that a strong pinning mechanism exists for the surface Fermi level and that there exist two well-defined surface transport channels with different mobilities and thicknesses. Considering that ARPES studies consistently show that downward band bending develops on Bi$_2$Se$_3$ surfaces (see Fig. 2(f)) [21-24], not only the topological surface states but also the two-dimensional electron gas (2DEG) states in the quantum confined accumulation layers can be the sources of these surface channels. If we assume that only the lowest level of the 2DEG is filled (see Supplemental Material SD [18] regarding this assumption), the sheet carrier densities ($n_{SC}$) of the topological surface state and the 2DEG should be given by $n_{SC,TI} = k_{F,TI}^2/(4\pi)$ and $n_{SC,2DEG} = k_{F,2DEG}^2/(2\pi)$, respectively, where $k_F$'s stand for the Fermi wave vectors and the factor of two difference is due to spin-degeneracy. Because $k_{F,2DEG} < k_{F,TI}$, we should always have $n_{SC,2DEG} < 2n_{SC,TI}$. Therefore, with $n_{SC-1}$ of ~$3.0 \times 10^{13}$ cm$^{-2}$ and $n_{SC-2}$ of ~$8 \times 10^{12}$ cm$^{-2}$, the inequality is satisfied only if $n_{SC-1}$ is from the TI band and $n_{SC-2}$ from the 2DEG, but not the other way around. Moreover, when the thickness of the sample approaches that of the 2DEG, the confinement will start affecting the energy levels of the 2DEG. Because the film is confined by air on one side and sapphire substrate on the other, the thickness confinement can be well approximated by the simple infinite square potential well model. For the infinite well, the lowest energy level from the bottom of the conduction band is given by $h^2/(8m^*t^2)$, where h is the Planck constant, m* is the effective electron mass and t is the film thickness. With $m^* = 0.15 m_e$ [22], where $m_e$ is the bare electron mass, this level is found to be 0.04 eV for t = 8 nm and 0.6 eV for t = 2 nm. When compared with the typical band-bending energy of 0.1~0.3 eV reported in ARPES studies [21-24], the 2DEG will start feeling the thickness effect by ~8 QL and will be severely affected by 2 QL. These analyses suggest that $n_{SC-1}$, which is constant down to 2 QL, is unlikely to originate from the 2DEG, whereas $n_{SC-2}$, which starts to change at ~8QL is more consistent with the expected behaviour of a 2DEG. ; see Supplemental Material SD and SE for further discussion [18].

According to the standard TI theory, the thickness of a topological surface state [25] is ~ 1 nm, which is given by $\hbar v_F/E_g$, where $v_F$ (= $4.5 \times 10^5$ m/s) is the Fermi velocity of the Bi$_2$Se$_3$ surface band and $E_g$ (= 0.3 eV) is the bulk band gap of Bi$_2$Se$_3$ [8, 26]; this implies that the thickness of the top and bottom surface states combined should be ~2 nm, which turns out to be exactly the thickness of the first channel, $n_{SC-1}$. If we assume that each of the top and bottom surfaces contribute equally to the observed carrier density of ~$3.0 \times 10^{13}$ cm$^{-2}$, $n_{SC,TI} = k_{F,TI}^2/(4\pi)$ provides $k_{F,TI}$ of 0.14 Å$^{-1}$, and this value is within the range that ARPES reports on band-bent Bi$_2$Se$_3$ samples [21-24]. There is a subtle point to discuss here, though. It is known from an ARPES study that the Dirac point on the surface band disappears for films thinner than 6 QL [27]. However, if the surface Fermi level is far from the Dirac point as depicted in Fig. 2(g), the sheet carrier density, which is simply a measure of $k_F^2$, should not be much affected by gap-opening at the Dirac point. Therefore, our observation of constant sheet carrier density of the topological surface states down to 2 QL is not in contradiction with this gap opening phenomenon at the Dirac point.

According to the above discussion, $n_{SC-2}$ is likely from an ~8 nm thick surface 2DEG. If we assume symmetric band bending on both the top and bottom surfaces, the 2DEG corresponds to $n_{SC,2DEG}$ of ~$4 \times 10^{12}$ cm$^{-2}$ on each surface with half the thickness. With $n_{SC,2DEG} = k_{F,2DEG}^2/(2\pi)$, this converts to $k_{F,2DEG}$ of 0.05 Å$^{-1}$. Interestingly, this Fermi wave vector is close to those obtained from Shubnikov-de Haas (SdH) oscillations of these samples; see Supplemental Material SE for the details [18]. However, no SdH oscillations were observed around $k_F$ of 0.14 Å$^{-1}$, the value associated with the topological surface band.

The mobilities in Fig. 2(e) also show thickness independence, within some error bars, down to 4~8 QL. However, unlike $n_{SC-1}$, which remained constant down to 2 QL, its mobility, $\mu_{SC-1}$, clearly degraded for ultrathin films. This difference can be understood by the fact that unlike the carrier density, which is simply a measure of the Fermi surface area, mobility is a measure of scattering time and thus susceptible to disorders and interactions, which is likely to become more significant for ultrathin samples. Another notable feature is that $\mu_{SC-1}$ is substantially smaller than $\mu_{SC-2}$ over the entire thickness range. This observation may look puzzling according to the common expectation that the mobility of the topological surface band should be high due to absence of backscattering. However, this expectation should be taken with caution. First of all, the topological protection mechanism guarantees only the metallicity of the surface state, and the mobility should still depend on the details of interactions. Considering that backscattering accounts for only a small fraction of the scattering [4] and that the topological surface state is spatially more confined than the 2DEG, there is no fundamental reason that the topological surface state should have a higher mobility than the 2DEG. Because high mobility ($\mu \gg 1/B$) is critical for the observation of SdH oscillations, we may or may not observe SdH oscillations from any of these surface channels, even if both are metallic. With $\mu_{SC-1} \approx 0.05$ m$^2$V$^{-1}$s$^{-1}$, $\mu_{SC-2} \approx 0.3$ m$^2$V$^{-1}$s$^{-1}$ and $B_{max} = 9$ T, we get $\mu_{SC-1}B_{max} \approx 0.5$ and $\mu_{SC-2}B_{max} \approx 3$. According to these numbers, we expect some SdH oscillations from SC-2 (2DEG) but none from SC-1 (topological surface band), and this expectation is experimentally confirmed in Supplemental Material SE [18].

Figure 3 presents another set of thickness-independent transport properties. In the normalized resistance vs. magnetic field data in Fig. 3(a), the cusp around zero magnetic field is an indication of the weak anti-localization (WAL) effect. Although Fig. 3(a) gives the impression that the magneto-transport is highly thickness-dependent, the small magnetic field regime in Fig. 3(b) provides a surprisingly simple picture. On the surface of TI materials, backscattering is at the minimum due to time-reversal symmetry when magnetic field is absent. With increasing magnetic field, which breaks the time-reversal symmetry, backscattering increases and leads to a reduction in conductance as in Fig. 3(b); this phenomenon is called the WAL effect [13-14]. Just like the other transport properties, this WAL effect also shows thickness independence for films thicker than ~8 QL. According to the standard WAL theory [28], the 2D magneto-conductance, G(B), is expected to change as $\Delta G(B) = A(e^2/h)[\ln(B_\phi/B) - \Psi(1/2 + B_\phi/B)]$ where A is a coefficient predicted to be $1/(2\pi)$ for each 2D channel, $B_\phi$ is the de-phasing magnetic field, and $\Psi(x)$ is the digamma function. The de-phasing magnetic field is related to the phase coherence length $l_\phi$ via $B_\phi = \hbar/(4el_\phi^2)$ [13-14].

Figure 3(c) shows that A remains almost constant from 3 through 128 QL, with a value between $1/(2\pi)$ and $1/\pi$. If the top and bottom surfaces were completely decoupled from each other with an insulating bulk state, A should be close to $1/\pi$. On the other hand, if the bulk of the film dominates and/or the bulk and two surfaces behave as a strongly coupled single entity, then the value should reduce to $1/(2\pi)$ [13]. Figure 3(c) shows that our films are somewhere between these two extremes. However, if the bulk contribution to the WAL effect were significant, $l_\phi$ should grow with thickness [13]. Therefore, $l_\phi$ being almost thickness-independent between ~8 and 128 QL in Fig. 3(d) is a clear indication that the observed WAL effect originates mainly from surface channels [13].

In summary, significant advances in $Bi_2Se_3$ thin film qualities allowed observation of dominant, thickness-independent surface transport channels. Conductance, sheet carrier densities, mobilities and WAL parameters remained nearly independent of thickness over two orders of thickness range. Such thickness-independent transport properties, trivially expected in TIs, were never observed before because of non-trivial bulk effects. In order to explain the observed surface transport properties, not only the topological surface states but also the quantum confined 2DEG channels have to be considered. How each of these different surface channels responds against various excitations is an important scientific/technological question that needs to be further investigated in future studies.

We thank Keun Hyuk Ahn, Peter Armitage, Eva Andrei, Liang Fu, and David Vanderbilt for discussions and comments. This work is supported by IAMDN of Rutgers University, National Science Foundation (NSF DMR-0845464) and Office of Naval Research (ONR N000140910749).

**Figure Legends**

**FIG. 1 (color online). Molecular beam epitaxy growth of $Bi_2Se_3$ films.**

(a) RHEED pattern of a typical $Bi_2Se_3$ film grown on an $Al_2O_3$ (0001) substrate by MBE. The sharp streaky pattern accompanied with the bright specular spot and Kikuchi lines is indicative of a high quality single crystalline growth. (b) 1.5 × 1.5 µm$^2$ scanned AFM image of a 300 QL thick $Bi_2Se_3$ film grown on $Al_2O_3$ (0001). Large terraces (largest ever reported for $Bi_2Se_3$ thin films) are observed, further verifying the high quality of the grown films.

**FIG. 2 (color online). Transport properties of $Bi_2Se_3$ films**

(a) Resistance vs. temperature for each thickness. (b) Conductance at 1.5 K as a function of thickness. (c) Hall resistance vs. magnetic field for a 16 QL sample plotted together with the two-carrier model fitting curve described in the text. (d) and (e) Sheet carrier densities and mobilities vs. thickness. For 2 and 3 QL films (shown by a diamond in the inset), the sheet carrier density was directly read off from the linear $R_{xy}$ vs B curve. In (b), (d) and (e), the horizontal straight lines are guides for illustration, and data for films thinner than 16 QL are plotted in the insets. (f) Conduction band minimum ($CB_{min}$) and valence band maximum ($VB_{max}$) along the depth of the sample, showing the downward band-bending toward the surface. (g) Schematic surface band diagrams, depicting how the surface bands change through the critical thickness (6 QL) when the surface Fermi level is high: CB and SS stands for the bulk conduction band and the topological surface state, respectively.

**FIG. 3 (color online). Weak anti-localization effect.**

(a) Normalized resistance change as a function of magnetic field, measured at 1.5 K, where $\Delta R(B) \equiv R(B) - R(0)$. Deep cusp in low field regime is characteristic of the WAL effect. (b) Conductance change vs. magnetic field in the low field regime: 8 - 128 QL curves are almost overlapping. The theoretical WAL fitting curves are plotted together for each data set. (c) and (d) The WAL fitting parameters, A and $l_\phi$ versus thickness, respectively. The horizontal lines are a guide for illustration.

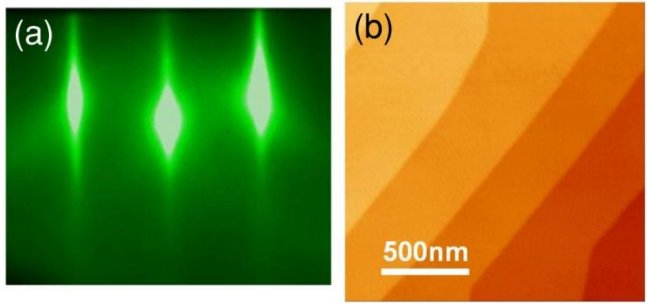

**Fig. 1 (One column width)**

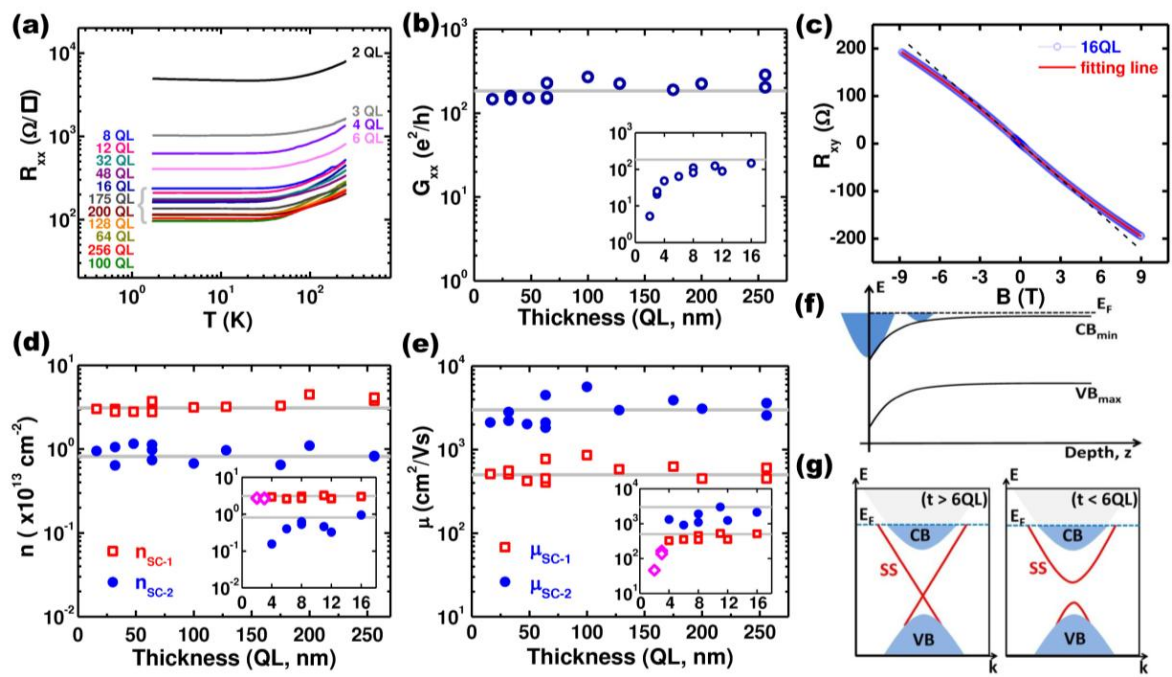

**Fig. 2 (Two column width)**

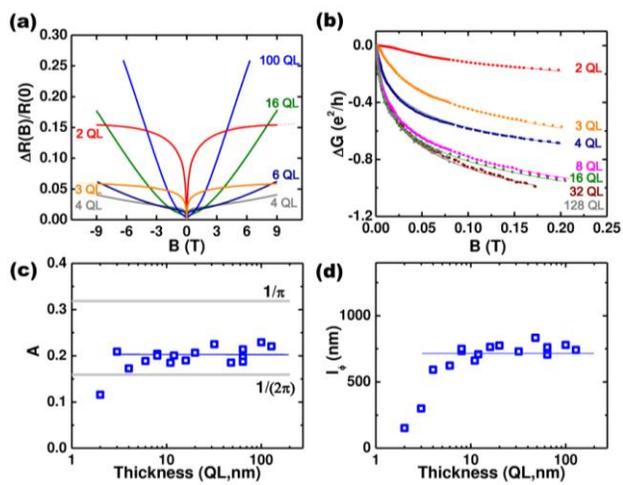

**Fig. 3 (One column width)**

Supplemental Material

# Thickness-independent transport channels in topological insulator $Bi_2Se_3$ thin films

Namrata Bansal, Yong Seung Kim, Matthew Brahlek, Eliav Edrey, and Seongshik Oh

## SA. Transport measurements

The transport measurements were carried out in an AMI superconducting magnet with a base temperature of 1.5K, and a maximum field of 9T. Resistance measurements were done with a Keithley 2400 Source Meter in conjunction with a Keithley 7001 Switch System. The $Al_2O_3$ substrates were 1 cm $\times$ 1 cm square; this geometry enabled us to use the standard 4-point van der Pauw method to measure resistance $R_{xx}$, and $R_{xy}$. We used thin indium wires (0.1 ~ 0.2 mm in diameter) to make contact with the sample at the corners, and a numerical symmetrisation procedure to eliminate unwanted mixing of $R_{xx}$ and $R_{xy}$ during the measurement process. All measurements were made within 20 minutes of the sample being taken out of the MBE chamber in order to minimize environmental factors.

## SB. Growth of $Bi_2Se_3$ films

High-quality $Bi_2Se_3$ films were grown on $Al_2O_3$ (0001) substrate in a custom-designed SVTA MOS-V-2 MBE system [S1]; the base pressure of the system was lower than $5 \times 10^{-10}$ Torr. Bi and Se fluxes were provided from Knudsen cells; the fluxes were measured using a quartz crystal microbalance, Inficon BDS-250, XTC/3.

To start with a clean $Al_2O_3$ (0001) (sapphire) substrate surface, we exposed the substrate to an *ex situ* UV ozone cleaning step before mounting it in the growth chamber to burn off majority of the organic compounds that may be present on the surface. To further remove any possible contaminants from the substrate surface, the sapphire substrate was heated to 700 $^o$C in oxygen pressure of $10^{-6}$ Torr for 10 min [S2]. The substrate surface was observed with RHEED before and after the treatment, and a bright specular spot and Kikuchi lines were observed after heating and then cooling the substrate. Figure S1a-b indicates that this procedure helped improving the surface conditions. $Bi_2Se_3$ films of various thicknesses were then grown using the two-temperature growth process[S1]. Evolution of the film surface during growth was monitored by RHEED, shown in Fig. S1(c-f). After deposition of 3 QL of $Bi_2Se_3$ at 110 $^o$C, a sharp streaky pattern was observed, indicating single-crystal $Bi_2Se_3$ structure. The film was then slowly annealed to a temperature of 220 $^o$C, which helped further crystallization of the film as seen by the brightening of the specular spot. The diffraction pattern and the Kikuchi lines became increasingly sharp on further $Bi_2Se_3$ deposition. This shows that the grown films have atomically flat morphology and high crystallinity. The film quality was further improved by annealing the sample at 220$^o$C for an hour after the growth. This process led to high quality single crystalline films with large terraces and minimal bulk conduction as described in the main text.

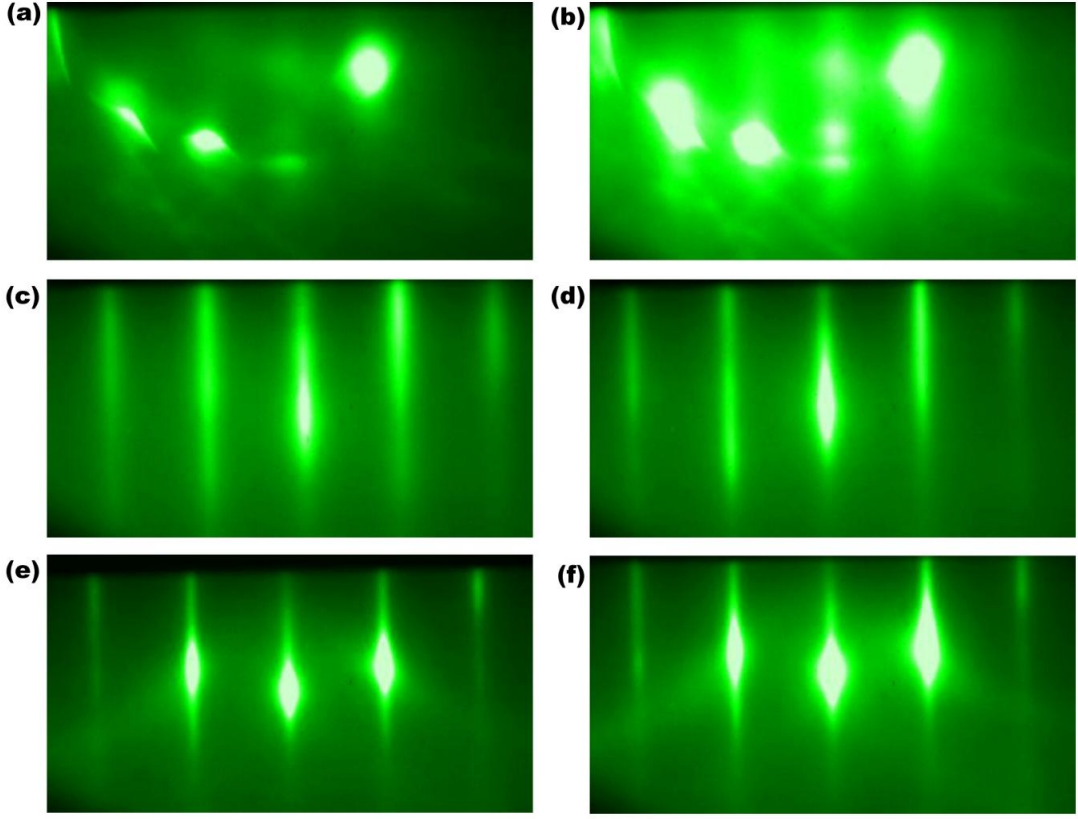

**Figure S1: RHEED images showing the steps of Bi$_2$Se$_3$ growth on sapphire substrates.** (a) Sapphire substrate mounted in the UHV growth chamber after UV-cleaned for 5 min. (b) On heating to 700 °C in an O$_2$ pressure of 1x10$^{-6}$ Torr for 10 min. (c) After deposition of 3 QL of Bi$_2$Se$_3$ film at 110 °C. (d) Specular beam spot gets brighter on annealing the film to 220 °C. (e) RHEED pattern gets much brighter and sharper on subsequent growth of another 29 QL at 220 °C. (f) Final RHEED pattern of the 32 QL film after being annealed at 220 °C for an hour.

## SC. 'G$_{xy}$ fitting' based on two-carrier model

With non-interacting, relaxation time approximation, under perpendicular magnetic field, the elements of the conductance tensor as a function of the two mobilities (µ$_1$, µ$_2$) and the corresponding sheet carrier densities (n$_1$, n$_2$) are given as [S3]:

$$G_{xx} = e\left(\frac{n_1\mu_1}{1+\mu_1^2 B^2} + \frac{n_2\mu_2}{1+\mu_2^2 B^2}\right); G_{xy} = eB\left(\frac{n_1\mu_1^2}{1+\mu_1^2 B^2} + \frac{n_2\mu_2^2}{1+\mu_2^2 B^2}\right).$$

Inverting this conductance tensor, we find the Hall resistance to be:

$$R_{xy}(B) = -\frac{G_{xy}}{G_{xx}^2 + G_{xy}^2} = -\frac{B}{e}\frac{(n_1\mu_1^2 + n_2\mu_2^2) + B^2\mu_1^2\mu_2^2(n_1+n_2)}{(n_1\mu_1 + n_2\mu_2)^2 + B^2\mu_1^2\mu_2^2(n_1+n_2)^2}$$

Although this formula can be used to fit the measured Hall resistance data as shown in Fig. 2c, we found that $G_{xy}(B)$ provides simpler and more reliable fitting results. Using $G_{xy}(B)$, we can easily reduce the number of fitting parameters from four to two as shown below.

If we take the limiting (B→0) case of the $G_{xx}(B)$ and $G_{xy}(B)$ expressions, they reduce to:

$$\frac{G_{xx}(0)}{e} = n_1\mu_1 + n_2\mu_2 = C_1; \quad \lim_{B\to 0}\frac{G_{xy}(B)}{eB} = n_1\mu_1^2 + n_2\mu_2^2 = C_2,$$

where e is the electronic charge and $C_1$ and $C_2$ are constants that can be found directly from the measured data. $C_1 = \frac{G_{xx}(0)}{e}$ can be directly read off from the measured conductance data at zero field, and $C_2 = \lim_{B\to 0}\frac{G_{xy}(B)}{eB}$ can be found from the linear slope of the Hall conductance near zero magnetic field. Solving the above two equations, we find:

$$n_1 = \frac{C_1\mu_2 - C_2}{\mu_1\mu_2 - \mu_1^2}; \quad n_2 = \frac{C_1\mu_1 - C_2}{\mu_1\mu_2 - \mu_2^2}.$$

This way, we can eliminate the two parameters, $n_1$ and $n_2$, from the Hall conductance equation such that:

$$G_{xy}(B) = eB\left(\frac{C_1\mu_1 - C_2}{(\mu_1/\mu_2 - 1)(1+\mu_2^2 B^2)} + \frac{C_1\mu_2 - C_2}{(\mu_2/\mu_1 - 1)(1+\mu_1^2 B^2)}\right).$$

Because this is now just two-parameter fitting, fitting can be much more reliably performed than the original four parameter fitting. This fitting provides two mobilities for each sample and their corresponding sheet carrier densities are then calculated from the mobilities.

Because it is the resistances not the conductances that are directly recorded from instruments, in order to obtain Hall conductance data from the measured resistance values, matrix inversion has to be done. In other words, the elements of conductance tensors are found from measured $R_{xx}$ and $R_{xy}$ as:

$$G_{xy}(B) = -\frac{R_{xy}}{R_{xy}^2 + R_{xx}^2}; \quad G_{xx}(B) = \frac{R_{xx}}{R_{xy}^2 + R_{xx}^2}.$$

These $G_{xy}$ data are in good agreement with the two-carrier model, as shown Fig. S2.

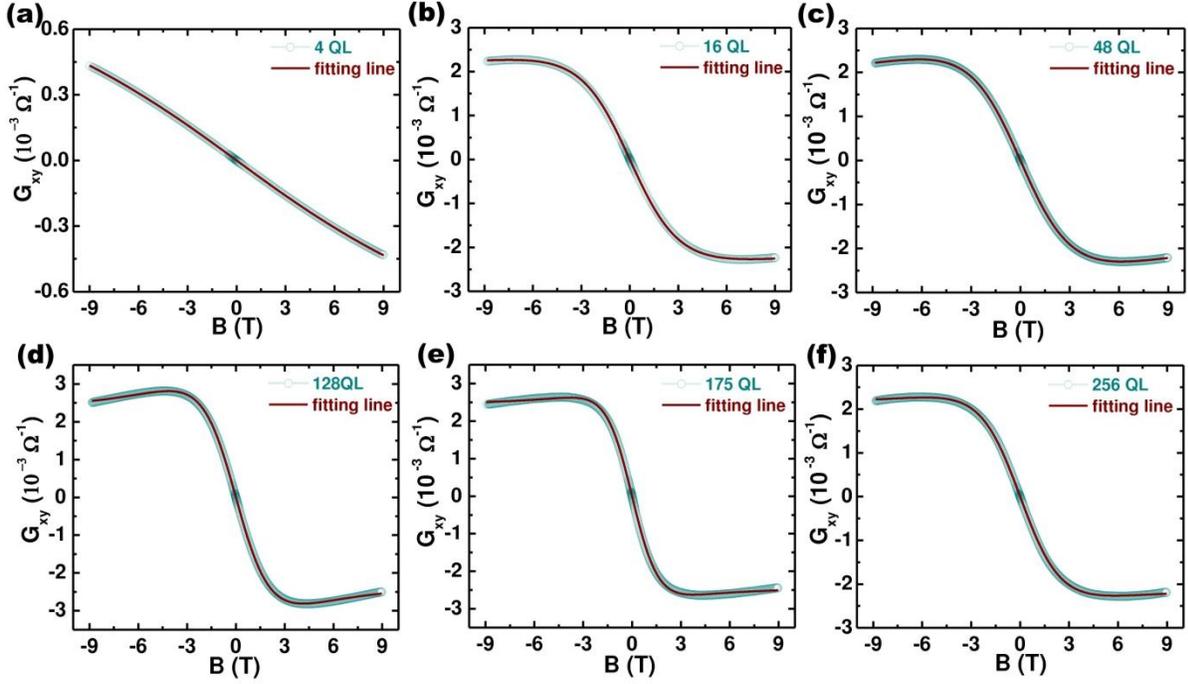

**Figure S2: Hall conductance data fitted with the standard two-carrier model.** In these plots, $G_{xy}(B) \equiv \frac{R_{xy}}{R_{xy}^2 + R_{xx}^2}$ was used; note the absence of the '–' sign.

## SD. Quantum confinement

ARPES studies consistently show that $Bi_2Se_3$ has a strong tendency to make downward band bending over time: in several hours in vacuum and seconds in air [S4-S7]. This downward band bending tends to develop 2DEGs on $Bi_2Se_3$ surfaces, and so it is important to consider both the 2DEGs and the TI surface states (SS) in transport studies.

As seen in the main text, the sheet carrier densities corresponding to the two conducting channels remain constant up to a film thickness of 256 QL. Then the bulk contribution excluding the 2DEG channel must be lower than either of these two channels. From $n_{SC-2} \sim 1 \times 10^{13}$ cm$^{-2}$ for the 256 QL sample, we get $n_{bulk} < n_{SC-2}/t \Rightarrow n_{bulk} < 5 \times 10^{17}$ cm$^{-3}$. Now from $n_{bulk} = k_{F,bulk}^3/(3\pi^2)$, where $k_{F,bulk}$ is the Fermi wave number, we get $k_{F,bulk} < 0.025$ Å$^{-1}$. Using effective mass $m^* \sim 0.15 m_e$ [S4], this limits the bulk Fermi level ($E_F$) to less than 15 meV from the bottom of the conduction band, $CB_{min}$.

According to our two-carrier model, if the surface states contribute equally on both surfaces, the spin non-degenerate carrier density due to SS is $n_{TI} \sim 1.5 \times 10^{13}$ cm$^{-2}$ giving a Fermi wave number, $k_{F,TI} \sim 0.14$ Å$^{-1}$ [$n_{TI} = k_{F,TI}^2/(4\pi)$]. On the other hand, the spin degenerate sheet carrier density for the 2DEG is given by $n_{2DEG} \sim 4 \times 10^{12}$ cm$^{-2}$, corresponding to $k_{F,2DEG} \sim 0.05$ Å$^{-1}$ [$n_{2DEG} = k_{F,2DEG}^2/(2\pi)$], which implies, with $m^* \sim 0.15 m_e$, that $E_F$ should be ~60 meV from the bottom of the 2DEG level.

These energy levels are shown together in Fig. S3. The dotted line is the bulk conduction band projected onto the surface. In ARPES spectrum, both the conduction band features and the 2DEG states may coexist. Most ARPES studies in the literature showing the band-bending effect were done on samples with much higher carrier densities ($>> 5\times10^{17}$ cm$^{-3}$) than ours, and considering that the band-bending effect is more severe for low carrier samples, it is plausible to expect that our samples have more severe band bending than the commonly available ARPES spectra in the literature.

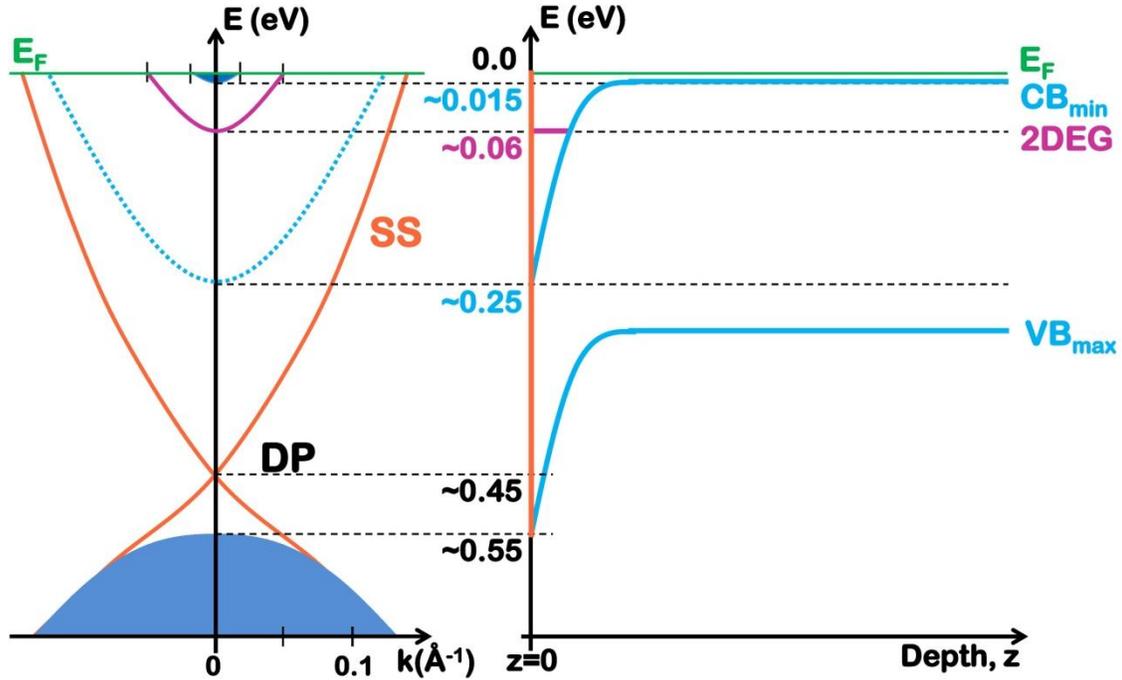

**Figure S3. Schematic of k-space energy dispersion and band-bending.** The left schematic shows the k-space energy dispersion on a Bi$_2$Se$_3$ surface. Without 2DEG, surface electrons will occupy all the way to the bottom of the conduction band, but if the quantum confinement works effectively, 2DEGs will form and there will be a gap between the minimum occupied level and CB$_{min}$ on the surface. Because the level of bend banding and quantum confinement depends on the bulk carrier density and the surface history, this schematic may not exactly match the ARPES spectra available in the literature. The right schematic describes the downward band bending near the surface.

The confinement along the z-direction (or the direction of growth) leads to quantized energy levels as shown in Fig. S4. As discussed in the main text, the carrier concentration corresponding to the accumulation layer increases slowly up to a thickness of ~8 QL and then remains almost constant, within an error bar, all the way up to 256 QL. Assuming a symmetric band bending at both interfaces, the 2DEG thickness can be taken as ~4 QL thick at each surface. In this scenario, for thick films (t $\gg$ 8 QL), the SS and the 2DEG on the opposite surfaces are well separated. As the film thickness is reduced, comparable to ~8 QL, the 2DEGs from opposite surfaces will start overlapping, and the thickness confinement will take over the band bending confinement as depicted in Fig. S4.

For an infinite well, the $n^{th}$ energy level from the bottom of the well (the conduction band minimum in this case) is given by: $E_n = (n\hbar\pi)^2/(2m^*t^2)$, where $\hbar$ is the reduced Planck's constant, $m^*$ is the effective electron mass (~0.15 $m_e$) and t is the thickness of the film. For a film thickness of 8 nm, the three lowest energy levels are found to be 0.04 eV, 0.16 eV, 0.35 eV, while the lowest energy level for a 2 nm thick films is 0.63 eV from the surface $CB_{min}$. Considering that the typical band-bending energy is 0.1~0.3 eV, as reported in various ARPES studies, as the film thickness is reduced below ~8 QL, the 2DEG levels will start rising, and for ultrathin films (2 or 3 QL), there will not be any allowed energy levels below the surface Fermi level, schematically shown in Fig. S4. This scenario is consistent with our observation of the two different types of metallic surface channels with negligible bulk contribution as described in the main text. Then the only conducting channel that can survive down to 2 QL is the TI surface states except that they have a gap opening at the Dirac point for less than 6 QL thick.

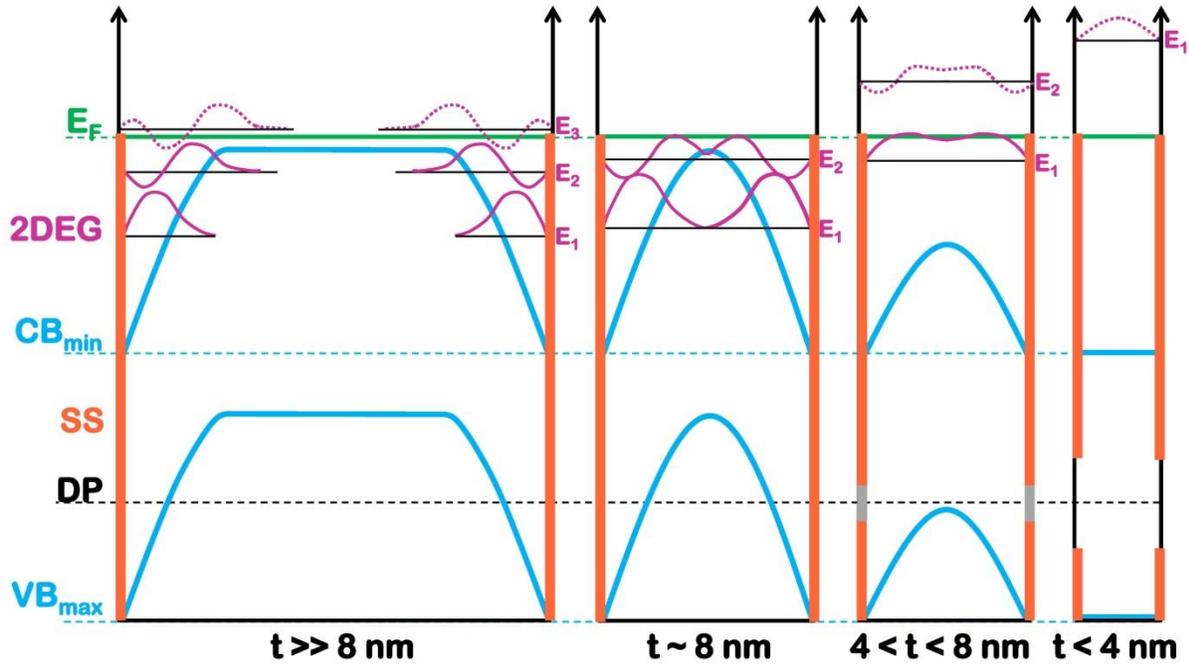

**Figure S4. Energy levels in the 2DEG.** This schematic shows how the quantized 2DEG levels in the accumulation layers are affected by the film thickness when the surface Fermi level is pinned. The approximate wavefunctions corresponding to the 2DEG energy levels are shown in purple; the surface states are in orange; $CB_{min}$ and $VB_{max}$ are shown in blue. For simplicity, energy level splittings due to wavefunction overlap are ignored in these schematics. Gap formation at the Dirac point (for t < 6 QL) is depicted as gray and black lines on the orange surface states.

## SE. SdH Oscillations

It is well known that Shubnikov-de Haas (SdH) effect is highly sensitive to the mobility of the sample and $\mu B \gg 1$ is required to observe clear SdH oscillations. From the Hall-effect measurements, the mobilities corresponding to the two conducting channels were found to be: $\mu_{SC-TI} \sim 0.05$ m$^2$V$^{-1}$s$^{-1}$, and $\mu_{SC\text{-}2DEG} \sim 0.3$ m$^2$V$^{-1}$s$^{-1}$. With the maximum field of 9 T ($\equiv$ 9 Vsm$^{-2}$), we get $\mu_{SC\text{-}TI}B_{max} \sim 0.5$ and $\mu_{SC\text{-}2DEG}B_{max} \sim 3$ and so the 2DEG state is more likely to exhibit the SdH oscillations than the surface state.

In fact, SdH oscillations were observed in many films with thickness of 16 QL or above. In all these samples, two SdH frequencies were observed (see Fig. S5(c-f)): $F_1 \sim 30$ T and $F_2 \sim 85$ T. These frequencies relate to the size of the Fermi surface via the Onsager relation, $F = (\hbar/(2\pi e))(\pi k_F^2)$, where $\hbar$ is the reduced Planck's constant, e is the electron charge and $k_F$ is the Fermi wave number. These numbers correspond to the wave vectors $k_{F1} \sim 0.03$ Å$^{-1}$ and $k_{F2} \sim 0.05$ Å$^{-1}$, respectively. If the Fermi surface is two dimensional, its 2D carrier density is related to $k_F$ as: $n_{2D} = k_F^2/2\pi$. The 2D carrier densities corresponding to these Fermi wave numbers are: $n_1 \sim 1 \times 10^{12}$ cm$^{-2}$ and $n_2 \sim 4 \times 10^{12}$ cm$^{-2}$; $n_{F1}$ and $n_{F2}$ plotted in Fig. S5(b) are twice these numbers taking into account top and bottom surfaces. As Fig. S5(b) shows, the sum of these two carrier densities match very nicely the 2DEG channel value we identified from the Hall measurement.

The very observation that the two very different measurements, the SdH oscillation and the two-carrier Hall effect fitting, provide very similar carrier densities for one of the Hall effect carriers strongly supports the reliability of our two-carrier model. As for the dominating SdH channel ($k_{F2} \sim 0.05$ Å$^{-1}$ with $n_2 \sim 4 \times 10^{12}$ cm$^{-2}$), if we assume a 3D Fermi surface with thickness-independent 3D carrier density, it is impossible to match with the Hall measurement, and so it definitely has to be of 2D nature, most likely a 2DEG as discussed above.

However, the story can be a little different for the smaller channel ($k_{F1} \sim 0.03$ Å$^{-1}$ with $n_1 \sim 1 \times 10^{12}$ cm$^{-2}$). Because the 2DEG is almost dominated by $k_{F2}$ channel, although $k_{F1}$ channel could also originate from another 2DEG level (say, $E_2$ from Fig. S4), we cannot completely rule out the possibility of it being a 3D channel. Based on the carrier density analysis of the thickest film (256 QL) in section SD, we estimated that $k_{F,bulk} < 0.025$ Å$^{-1}$, and so $k_{F1} \sim 0.03$ Å$^{-1}$ could be within this limit, considering the error bars in our measurement. Detailed angle dependent studies of each channel up to higher magnetic field will be needed to resolve this issue completely.

In summary, the channel with $k_F \approx 0.05$ Å$^{-1}$ observed in both SdH oscillation and Hall effect is best described by the 2DEG; and the smaller channel with $k_F \approx 0.03$ Å$^{-1}$ detected in SdH oscillation could be either a higher-level 2DEG or a residual bulk channel; but the channel with $k_F \approx 0.14$ Å$^{-1}$, which appeared only in Hall effect and maintained its metallicity all the way down to 2 QL, cannot be explained by the 2DEG and is most likely from the TI surface states.

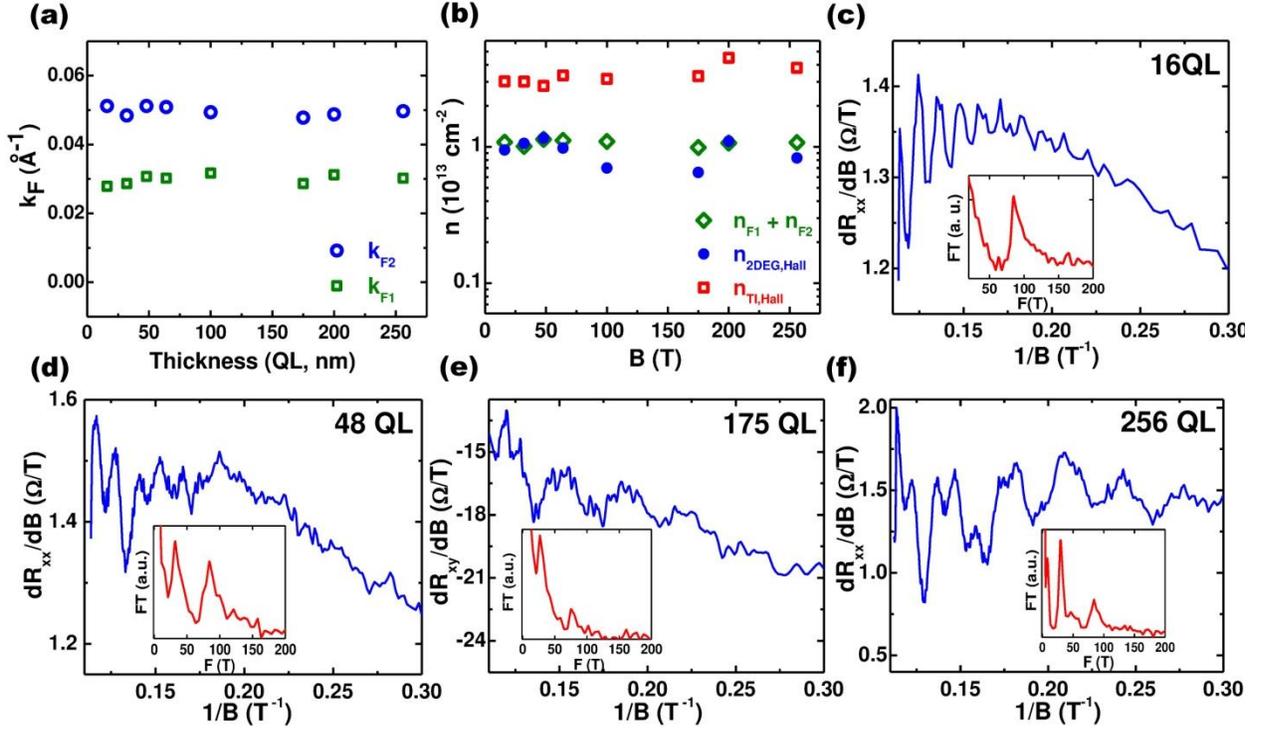

**Figure S5. SdH measurements. (a)** SdH-measured Fermi wave vectors as a function of sample thickness. **(b)** Comparison of the sheet carrier densities obtained from the two-carrier fitting of the Hall effect and the equivalent sheet carrier densities estimated from **(a)** as a function of thickness. **(c-f)** SdH oscillations observed in $dR_{xx}/dB$ and $dR_{xy}/dB$ for various thicknesses; corresponding Fourier transforms are shown in the insets.